\documentclass[aps,prd,onecolumn,preprintnumbers,nofootinbib,superscriptaddress,amsmath]{revtex4-2}
\usepackage[english]{babel}
\usepackage{bm}
\usepackage[utf8]{inputenc}
\usepackage[colorinlistoftodos, color=green!40, prependcaption]{todonotes}
\usepackage{amssymb}
\usepackage[pdftex, pdftitle={Article}, pdfauthor={Author}]{hyperref} 

\def\apjl{{Astrophysical Journal Letters}}             
\def\apj{{Astrophysical Journal}} 
             
\def\mnras{{Monthly Notices of the Royal Astronomical Society}}
 
\def\jcap{{Journal of Cosmology and Astroparticle Physics}}
\begin{document}
\title{Exploring cosmological constraints on galaxy formation time}
\author{Agripino Sousa-Neto }
\email{agripinoneto@on.br}
\affiliation{Observatório Nacional, Rio de Janeiro - RJ, 20921-400, Brasil}

\author{Maria Aldinêz Dantas}
\email{aldinezdantas@uern.br}
\affiliation{Departamento de Física, Universidade do Estado do Rio Grande do Norte, 59610-210, Mossoró - RN, Brasil }%

\author{Javier E. González}
\email{javiergonzalezs@academico.ufs.br}
\affiliation{Universidade Federal de Sergipe, São Cristóvão - SE, 49107-230, Brasil}

\author{Joel C. Carvalho}
\email{jccarvalho@on.br}
\affiliation{Observatório Nacional, Rio de Janeiro - RJ, 20921-400, Brasil}

\author{Jailson Alcaniz}
\email{alcaniz@on.br}
\affiliation{Observatório Nacional, Rio de Janeiro - RJ, 20921-400, Brasil}

\date{\today}

\begin{abstract}
The Universe consists of a variety of objects that formed at different epochs, leading to variations in the formation time which represents the time elapsed from the onset of structure formation until the formation time of a particular object. In this work, we present two approaches to reconstruct and constrain the galaxy formation time $t_f(z)$ using non-parametric reconstruction methods, such as Gaussian Processes (GP) and High-performance Symbolic Regression (SR). Our analysis uses age estimates of 32 old passive galaxies and the Pantheon+ type Ia supernova sample, and considers two different values of the Hubble constant $H_0$ from the SH0ES and Planck Collaborations. When adopting the $\Lambda$CDM model and the GP reconstructions, we find $\left<t_f\right>=0.72_{-0.16}^{+0.14}$ Gyr (SH0ES) and $\left<t_f\right>=1.26_{-0.11}^{+0.10}$ Gyr (Planck). Without considering a specific cosmological model, we obtain $\left<t_f\right>=0.71 \pm {0.19}$ Gyr (SH0ES) and $\left<t_f\right> = 1.35_{-0.23}^{+0.21}$ Gyr (Planck). Similar values are obtained from the SR reconstructions, with both methods (GP and SR) indicating the same behavior regarding the time evolution of $t_f(z)$. The  results also show significant differences in the formation time from SH0ES and Planck values,  highlighting the impact of the $H_0$ tension on the cosmological estimates of $t_f(z)$. In particular, the different approaches used in the analysis agree with each other, demonstrating the robustness and consistency of our results. Overall, this study suggests that galaxies have different evolutionary timescales and that $t_f$ is not constant, with noticeable variations at lower redshifts ($z \lesssim 0.5$).

\end{abstract}
\maketitle
\section{Introduction}

The formation of galaxies represents a central theme in cosmology, a process extensively investigated through both theoretical modeling and observational data. A key aspect of this research involves understanding the formation time ($t_f$), defined as the cosmic time required for a galaxy to assemble the bulk of its stellar mass. This timescale is crucial for constraining models of cosmic structure formation and for testing our current cosmological paradigm, consequently providing fundamental insights into galactic evolution.

Over the past few decades, our understanding of galaxy formation has advanced significantly, based on both observations and theoretical progress. Seminal works have profoundly shaped our knowledge of galaxy formation and evolution, including those by  \citep{Eggen1962,Peebles1965,Peebles1985,PressandSchechter1974,WhiteandRees1978,Blumenthal1984}. 
These foundational studies elucidated vital mechanisms governing the formation of cosmic structures and introduced key concepts such as hierarchical clustering, the decisive role of dark matter, and the importance of cooling processes in baryonic matter. They continue to be essential references for comprehending the timing and mechanisms of galaxy formation within the broader cosmological context.

In principle, galaxy formation does not necessarily occur on a universal timescale; instead, it may vary with redshift and environment. This study aims to investigate the $t_f$  redshift
dependence and  characterize its behavior throughout cosmic evolution. Here, we specifically focus on a particular class of galaxies, namely, passively evolving galaxies, like early-type galaxies (ETG). These objects exhibit little to no ongoing star formation, are dominated by older stellar populations that evolve slowly, and possess well-characterized colors, spectra, and luminosities. Their age can be calculated by modeling the stellar populations using either simple or composite stellar population models. Although our calculations rely on observational data of ETGs, it is important to acknowledge that the method employed is not entirely model-independent; the formation time can only be inferred from galaxy ages by assuming a cosmological model.

In the following sections, we will present two distinct approaches to constrain the redshift evolution of $t_f(z)$. These approaches involve the reconstruction of the age-redshift relationship and the luminosity distance, obtained through two non-parametric techniques: Gaussian Processes and High-Performance Symbolic Regression. In our analysis, we use age estimates from 32 old passive galaxies combined with data from the Pantheon+ Type Ia supernova sample. Our results suggest that galaxies exhibit different evolutionary timescales, indicating that $t_f(z)$ is not constant, with significant variations at lower redshifts ($z \lesssim 0.5$).

We organize this paper as follows: Our definition of formation time is presented in Section ~\ref{DGFTR}, while Section~\ref{ABLRGF} provides a brief review of the existing literature on formation time estimates from theory, simulations, and direct observation. Section IV details our methodology, including the data selection and reconstruction techniques used to constrain $t_f(z)$. Section~\ref{result} presents our results, while Section~\ref{conclu} summarizes our main conclusions.

\section{Galaxy formation time}\label{DGFTR}

In this work, we shall adopt the following definition for the galaxy formation time: $t_f$ represents the cosmic time, measured from $t = 0$, at which a given fraction of a galaxy's final stellar mass had formed. 
This is clearly a simplification that does not introduce significant error in the estimates $t_f$. As is well known, the process of galaxy formation commences only after a sufficient cosmic structure has emerged and the universe has cooled adequately. The gravitational seeds of galaxy formation are large-scale dark matter halos ($>10^8M_{\odot}$) that typically begin to collapse around 100–300 million years after the Big Bang, roughly corresponding to redshifts $z \sim 15-20$. 

If the age of a galaxy at a given redshift $z$ is denoted by $t_g$, representing the time elapsed since its stellar population was largely assembled (the galaxy's current age), then $t_g$ reflects the duration available for stellar evolution and other post-formation processes. The formation time can thus be expressed as
\begin{equation} \label{tf}
    t_f = t_u - t_g,
\end{equation}
where $t_u$ is the age of the Universe at redshift $z$. In the literature, related terms such as `formation epoch', `delay factor', or `incubation time' may be used interchangeably with $t_f$, depending on the specific modeling or observational context.


It is crucial to recognize that the above equations establish a direct equivalence between $t_f$ and the formation redshift ($z_f$). Consequently, for any given $z_f$ and a specified cosmological model, one can easily calculate $t_f$. This relationship is particularly important because numerous studies report observational measurements and estimates primarily in terms of formation redshift rather than formation time.

The theoretical age-redshift relation can be written as~\cite{Alcaniz:1999kr,Alcaniz:2001uy}:
\begin{equation}\label{eq:t_0}
t_u(z,\textbf{p})=H_0^{-1} \int_{z}^{\infty} \frac{dz'}{(1+z') \mathcal{H}(z',\textbf{p})} \;,
\end{equation}
where $\textbf{p}$ represents the set of cosmological parameters, $H_0=100h^{-1} \mathrm{kms^{-1}Mpc^{-1}}$ is the Hubble constant, 
and $\mathcal{H}(z', \textbf{p})$ is the dimensionless Hubble parameter normalized to its present-day value. This expression is essential for translating formation redshifts into cosmic times, allowing a comparison between theoretical predictions and observationally inferred ages of galaxies.

\section{A brief literature review on galaxy formation}\label{ABLRGF}

This section summarizes key findings from theoretical studies, simulations, and observations over the past decades, with a focus on estimating galaxy formation timescales — particularly formation time and redshift ($z_f$). Analytical models typically predict broad ranges rather than precise values. \citet{Fowler1987} estimated $t_f = 1.0 \pm 0.4$~Gyr after the Big Bang. \citet{Peebles1999} suggested star formation began at $z_f \sim 5$–6, while models such as fast clumpy collapse \cite{Thomas_1999} found $z_f \geq 2$ for cluster ellipticals. These theoretical results consistently suggest early formation epochs, especially for massive galaxies characterized by old stellar populations, such as Luminous Red Galaxies (LRGs).

Numerical simulations have reinforced these early-formation scenarios. \citet{Kauffmann1993, Baugh1998} showed that dwarf galaxy formation is suppressed between $z = 1.5$ and 5, with massive galaxies forming at $z_f \sim 3.5$. \citet{DeLucia2006}, using the Millennium Simulation, found lookback times of 8.6–11.4~Gyr for ellipticals, equivalent to $z_f \sim 5$. SPH simulations~\citep{Naab2007,Wellons2015} reported $z_f \sim 4$–5 for massive compact galaxies, with 50\% of their mass assembled by $z_f = 4.8$ while \citet{Yajima2022} studied galaxy formation in protocluster regions and reported assembly times from $z_f \sim 10$ for the first galaxies to $z_f \sim 3$ for high-mass systems, supporting an emphasis on environmental influences in $z_f$.

Observationally, early estimates based on halo collapse times~\cite{Eggen1962} and quasar metallicities~\cite{Peebles1985} placed galaxy formation at $z_f > 3$. Hubble Deep Field and spectroscopic surveys~\cite{Glazebrook2004,Daddi2005,Thomas2005} confirmed that many massive quiescent galaxies formed the bulk of their stars at $z_f > 2$, with formation timescales often below 1–2~Gyr. 
The star formation histories of over 22,000 SDSS early-types analyzed by \citet{Jimenez2007} revealed that galaxies with velocity dispersion $\sigma > 200$ km/s formed over 90\% of their stellar mass by $z_f > 2.5$, with star formation and chemical enrichment lasting $t_f \sim 1$–2 Gyr and quenching by $z > 1.2$.  \citet{Thomas2010} supported this result, placing $z_f \sim 4$ and lookback times of 12 Gyr for massive ellipticals in a study of 28,954 SDSS galaxies. 

Recent studies have strengthened the picture of ``downsizing,'' where more massive galaxies form earlier and faster. Spectral modeling~\cite{Pacifici2016,Carnall2019} revealed that high-mass quiescent galaxies at $z \sim 1$–2 typically formed at $z_f > 2.5$–4, with $t_f$ as short as 1~Gyr. JWST observations~\cite{Nanayakkara2024} pushed these limits further, identifying galaxies at $z \sim 3$–4 that formed at $z_f \sim 10$–20 within $t_f \sim 0.2$–0.5~Gyr.

A consistent observational trend has emerged: the formation redshift $z_f$ of quiescent galaxies increases with their observed redshift $z$. Studies~\cite{Pacifici2016,Carnall2018,Carnall2019} using spectroscopic samples and Bayesian SED fitting have demonstrated this correlation. This supports a downsizing scenario and motivates the use of new independent methods to constrain $t_f$ and $z_f$, contributing to a deeper understanding of quenching mechanisms and red sequence assembly.


\section{Methodology and Data}\label{method}

One of our primary goals is to demonstrate that it is possible to constrain the galaxy formation time using current cosmological data sets. To achieve this, we aim to develop a model-independent method to reconstruct galaxy ages and, consequently, $t_f$. 

We use Gaussian Processes (GP) and High-performance Symbolic Regression (PySR) to reconstruct galaxy ages. The GP method has been widely employed in cosmology for various applications, including null tests of the FLRW metric~\cite{2012SeikelPhRvD,2010Shafieloo}, reconstruction of cosmological matter perturbations \cite{2017Gonzalez,2016Gonzalez,2017Gonzalez_prd}, estimates of the Hubble constant \cite{2014Busti}, among others.
On the other hand, symbolic regression has recently been applied to investigate a potential dynamical behavior of dark energy, particularly in light of recent baryon acoustic oscillation (BAO) data from the DESI collaboration~\cite{Sousa-Neto:2025gpj}.

We follow two approaches to determine $t_f$. The first one calculates $t_f$ as the difference between the theoretical curve of the age of the Universe, obtained from Eq.~\eqref{eq:t_0}, and the galaxy age reconstruction via GP and PySR, where the theoretical curve is based on the $\Lambda$CDM model for two different values of $H_0$ reported by the Planck \cite{aghanim2020planck} and SH0ES \cite{Riess_2019} collaborations. The second approach uses a reconstruction of the lookback-time redshift relation from SNe data combined with galaxy ages, assuming that objects at the same redshift share the same age.

\subsection{Gaussian Processes}

Gaussian process is a model-independent method to smooth the data. Thus, this method allows one to reconstruct a function from data without assuming a parameterization for it. We use the code GP in Python (GaPP) to derive our GP results \cite{Seikel_2012}. The distribution of functions provided by GP is suitable for describing the observed data. At each point $z$, the reconstructed function $f(z)$ is also a Gaussian distribution with a mean value and Gaussian error. The functions at different points $z$ and $\tilde{z}$ are related by a covariance function $k(z,\tilde{z})$, which depends only on a set of hyperparameters $l$ and $\sigma_f$. Here, $l$ gives a measure of the coherence length of the correlation in the x-direction and $\sigma_f$ denotes the overall amplitude of the correlation in the y-direction. Both will be optimized by GaPP with the observed data set. In contrast to parametric methods, GP does not specify the form of the reconstructed function. Instead, it characterizes the typical changes in the function.

The different choices of the covariance function may affect the final reconstruction. In order to test the effect of this choice on the results, we perform reconstructions using the square exponential, Matérn ($\nu=$
3/2, 5/2, 7/2, 9/2) (see \citet{2013arXiv1311.6678S}) and Cauchy covariance functions and we do not obtain appreciable variations in the final results. Therefore, in what follows, we adopt the
squared exponential form \cite{Seikel_2012},
\begin{equation}
k(z,\tilde{z})=\sigma_f^2\exp\left({-\frac{(z-\tilde{z})^2}{2l^2}}\right)\;.
\end{equation}

\subsection{High-performance Symbolic Regression}

Symbolic Regression (SR) is a supervised learning task that focuses on identifying analytic expressions as models. It is typically approached through a multi-objective optimization framework, which aims to minimize both prediction error and model complexity simultaneously. Instead of fine-tuning specific parameters within an overly complex general model, SR explores the space of simple analytic expressions to find models that are both accurate and interpretable. Additionally, SR can be used as a nonparametric method, particularly for reconstructing cosmological observables.

An important aspect of SR is the choice of the set of operators, which are mathematical functions or symbols used to construct models that describe the relationship between input variables and predictions. In our analysis, we use the SR library called PySR~\cite{cranmer2023interpretablemachinelearningscience}. PySR is a powerful open-source library designed to efficiently discover symbolic models through a combination of evolutionary algorithms and gradient-based optimization. We adopt a set of unary operators — \{\texttt{log}, \texttt{sqrt}, \texttt{exp}, \texttt{cube}, \texttt{square}, \texttt{inv}\} — and binary operators — \{\texttt{+}, \texttt{-}, \texttt{*}, \texttt{/},\texttt{$^{\wedge} $}\} — as implemented in the PySR library.

Another crucial component of SR is the definition of the loss function, which is a key factor in optimizing the model's performance by measuring the difference between predicted and input values. We adopt the P-th power absolute distance loss\footnote{For more details, we refer the reader to   \url{https://ai.damtp.cam.ac.uk/symbolicregression/stable/losses/}}, defined as 
\begin{equation}
\mathcal{L} = |x - y|^P\;, 
\end{equation}
where $x$ represents the predicted value, $y$ denotes the true value, and $P$ is a positive real number that controls the sensitivity of the loss to errors. The methodology used with PySR follows the same procedure as in \cite{Sousa-Neto:2025gpj}, where further details can be found, such as the choice of the best-fit equation.

\begin{figure}[t]
    \centering
    \includegraphics[width=0.63\linewidth]{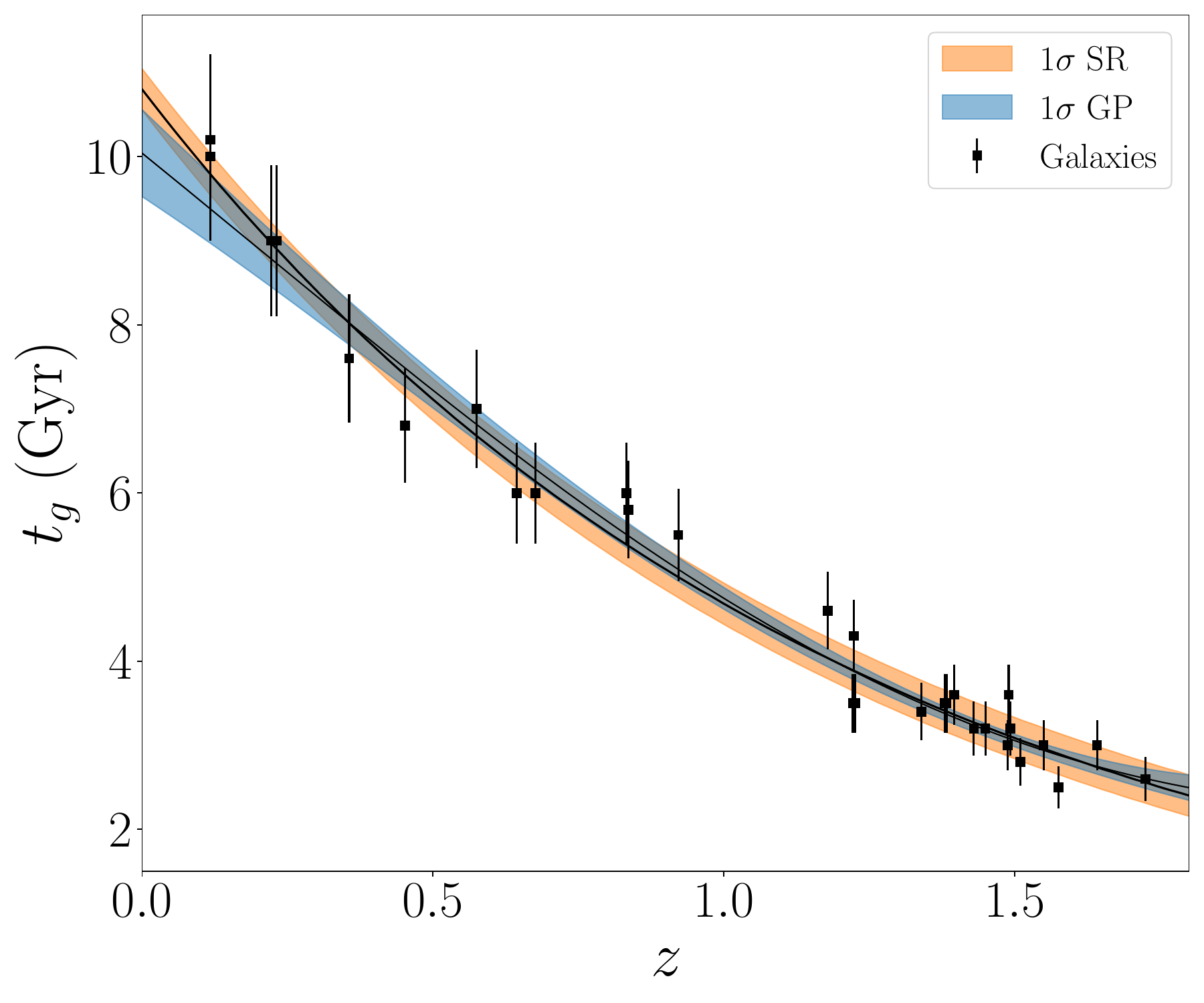}
    \caption{GP and SR reconstructions of the age-redshift relation from age estimates of 32 old passive galaxies (black points)~\cite{simon2005constraints}. The  contours show the reconstruction at $1\sigma$ level while the solid black line represents the mean of the reconstructions.}
    \label{fig:agegapp}
\end{figure}

\subsection{Age Data}

In this work, we use age estimates of 32 old passive galaxies distributed over the redshift interval $0.11 \leqslant  z \leqslant 1.84$, as analyzed by \citet{simon2005constraints}. The total sample is composed by three sub-samples:
10 field early-type galaxies \citep{1999Treu,2001Treu,2002Treu}, whose ages were obtained by using the SPEED models of  \citet{2004Jimenez}; 20 red galaxies from the publicly released Gemini Deep Deep Survey (GDDS) whose integrated light is fully dominated by evolved stars \citep{abraham2004gemini,McCarthy_2004}. Simon \textit{et al}. re-analyzed the GDDS old sample by using a different stellar population model and obtained ages within $0.1$ Gyr of the GDDS collaboration estimates and the two radio galaxies LBDS 53W091 and LBDS 53W069 \citep{1996Natur.381..581D,1997ApJ...484..581S}. The age data set used in this analysis is shown in Fig.~\ref{fig:agegapp}.

\begin{figure*}[t]
{\includegraphics[width=0.32\textwidth]{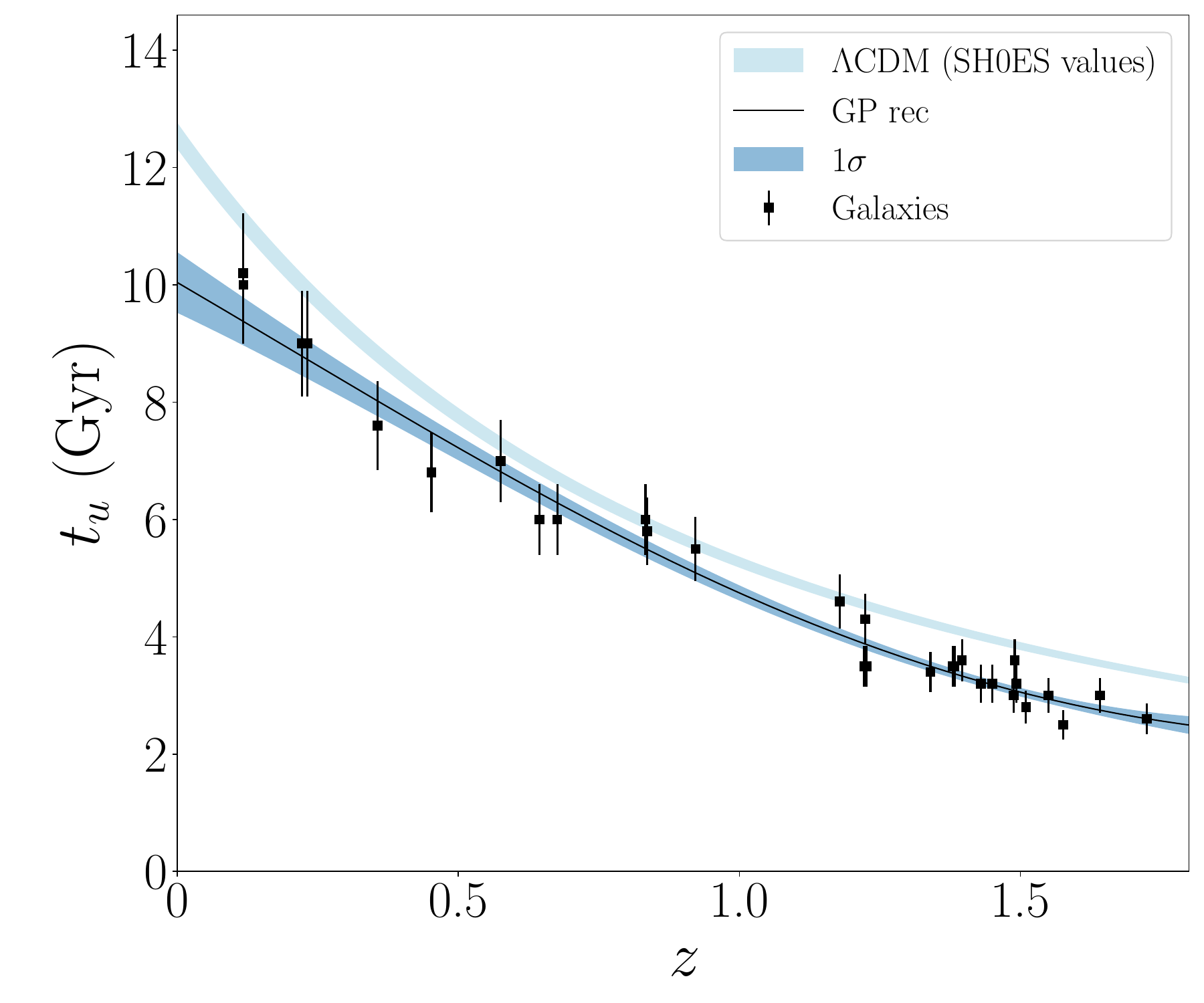}} 
{\includegraphics[width=0.32\textwidth]{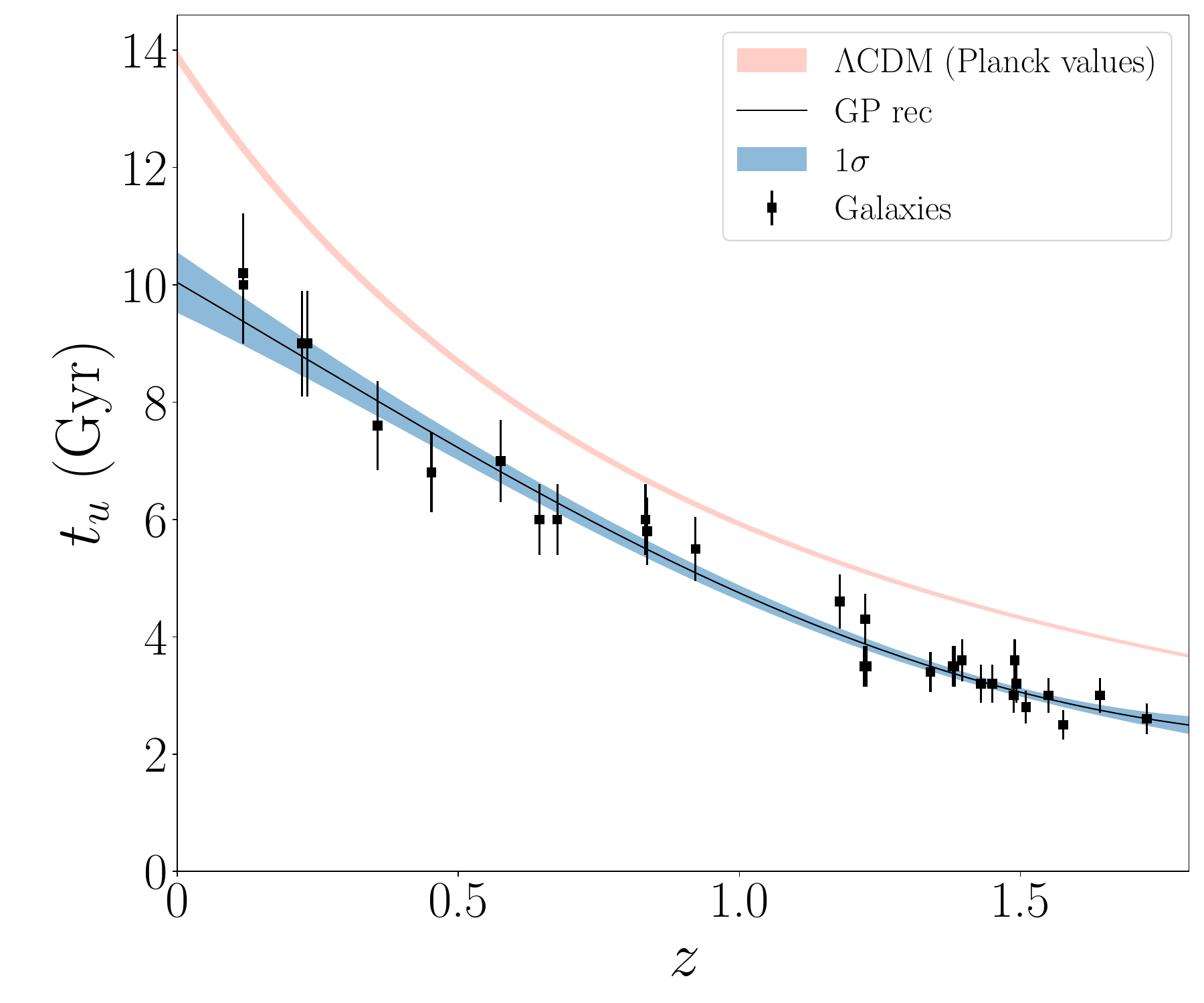}} 
{\includegraphics[width=0.32\linewidth]{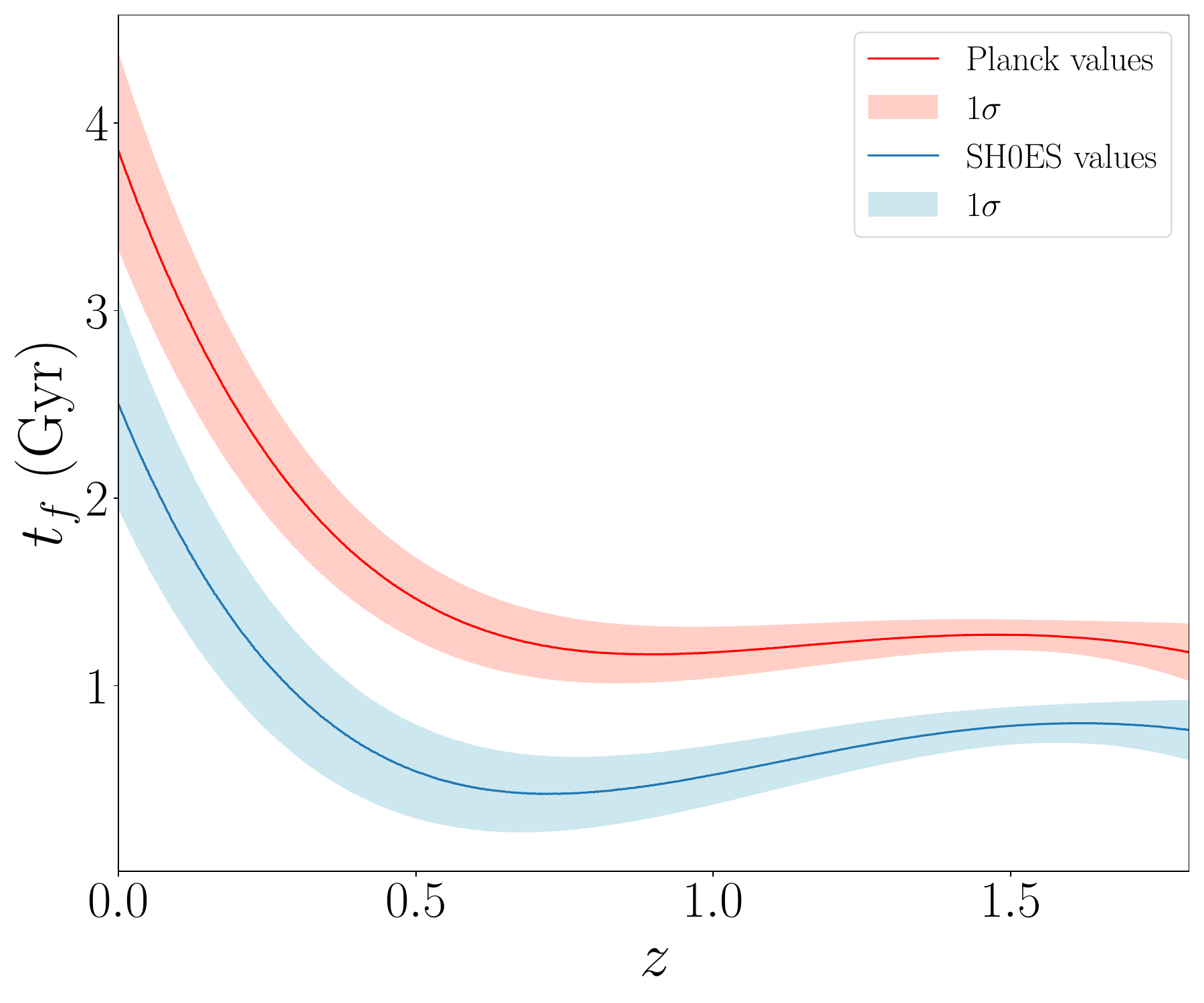}} 
\caption{\textit{Left}: The age-redshift relation assuming a flat, $\Lambda$CDM model and the  values of $H_0$ and $\Omega_m$ provided by the SH0ES collaboration. \textit{Middle}: The same as in the previous Panel for the values of $H_0$ and $\Omega_m$ provided by the Plank collaboration. In both panels, the dark blue region corresponds to the reconstructed age-redshift relation shown in Fig.~\ref{fig:agegapp}. \textit{Right}: Evolution of the galaxy formation time $t_f(z)$ with redshift obtained from the difference between $t_u$ and $t_g$ -- Eq.~\eqref{tf}. The red and blue regions represent the reconstructed $1\sigma$ intervals.}
\label{fig:30}
\end{figure*}

\begin{figure*}[t]
{\includegraphics[width=0.32\textwidth]{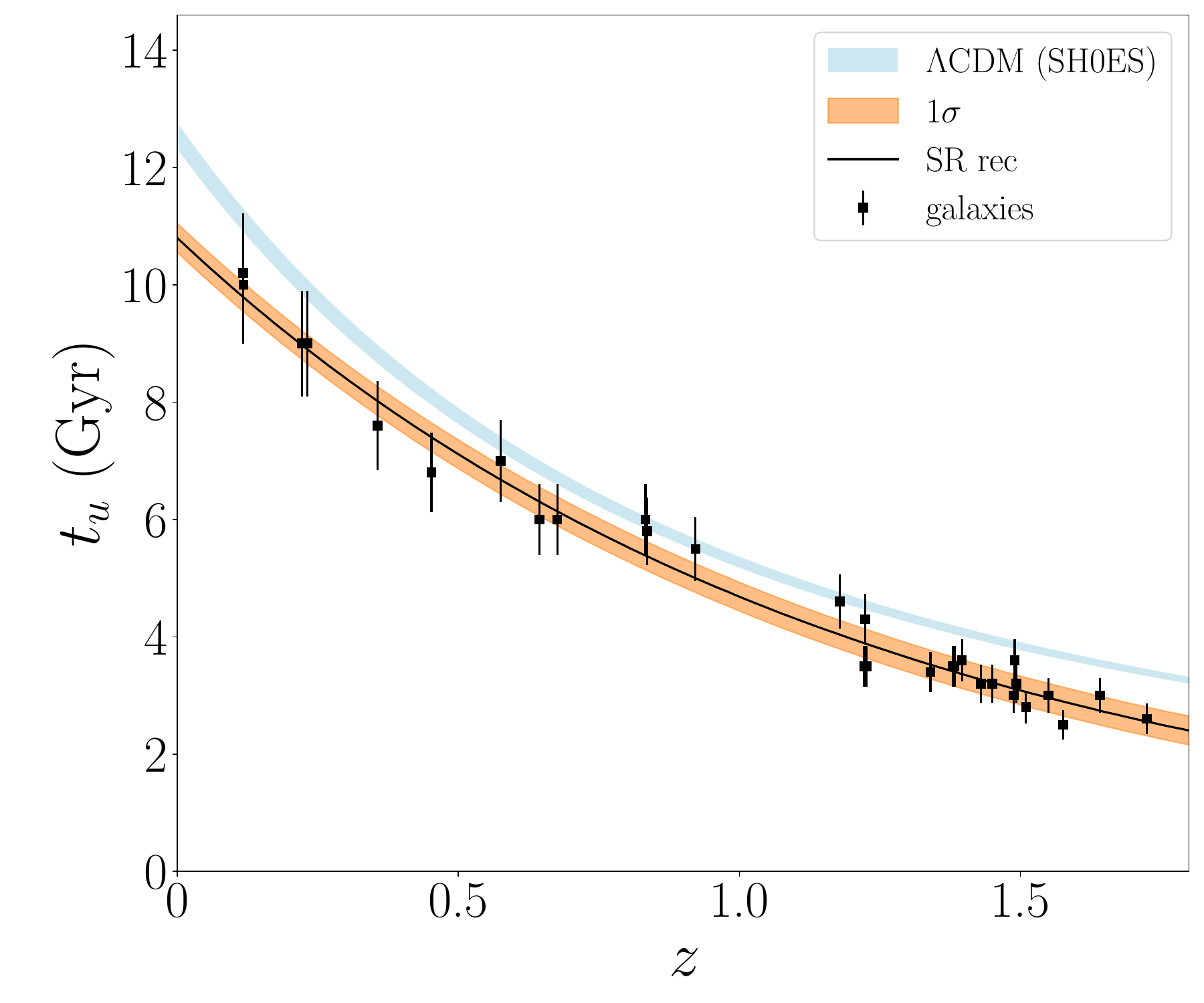}} 
{\includegraphics[width=0.32\textwidth]{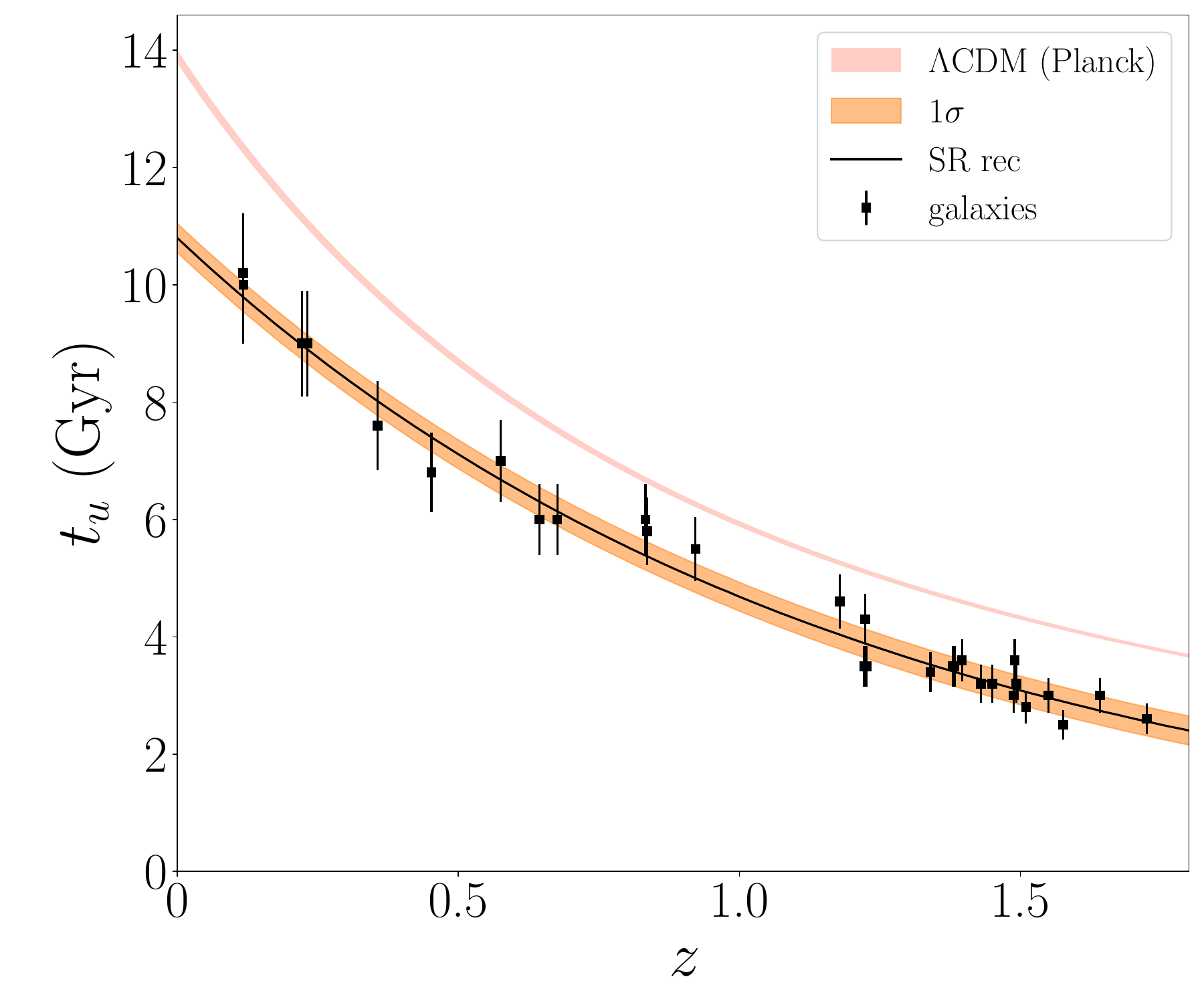}} 
{\includegraphics[width=0.32\linewidth]{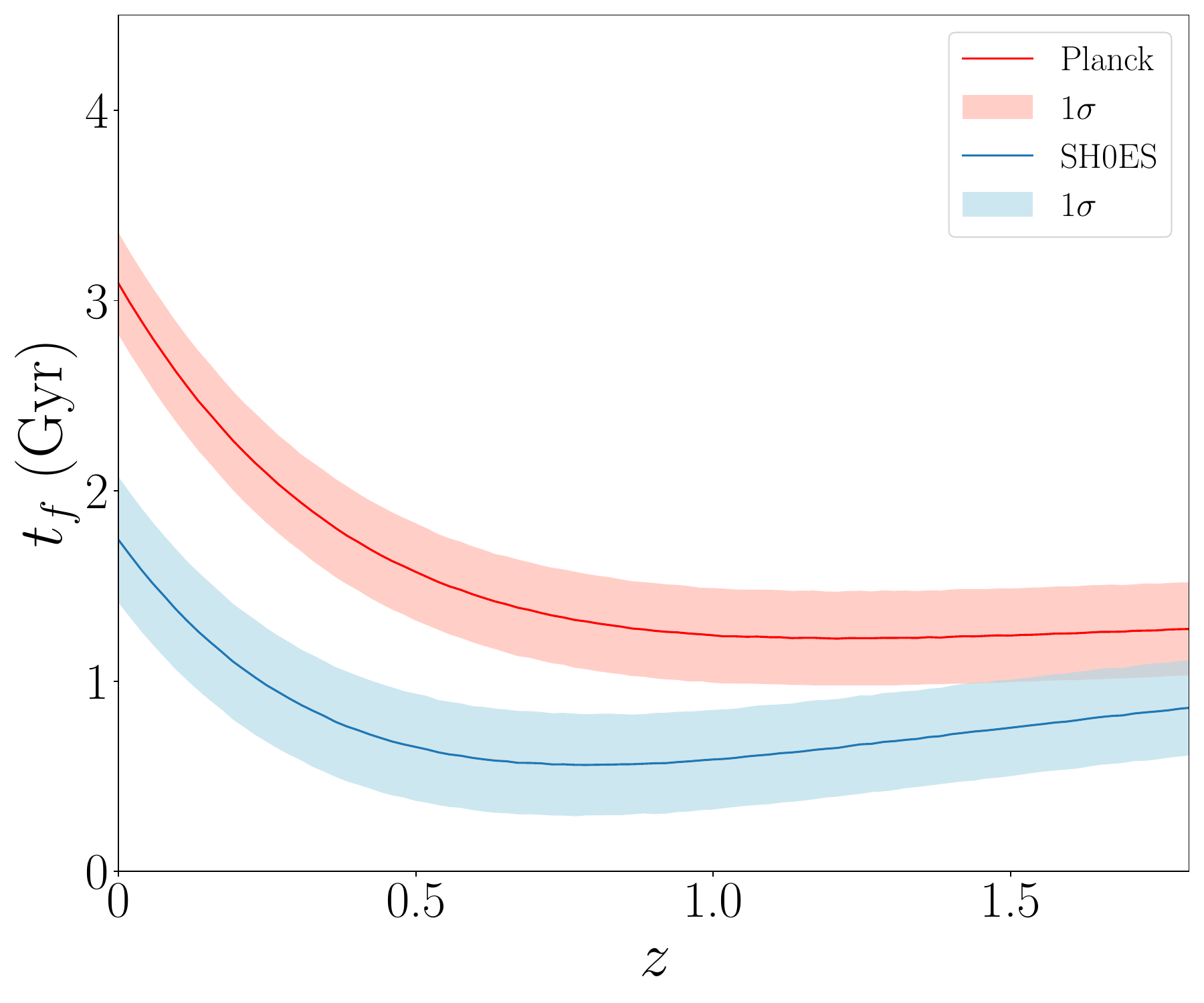}} 
\caption{Same as Fig.~\ref{fig:30}, but using PySR to reconstruct galaxy ages.}
\label{fig:pysr}
\end{figure*}

\begin{figure*}[t]
{\includegraphics[width=0.32\linewidth]{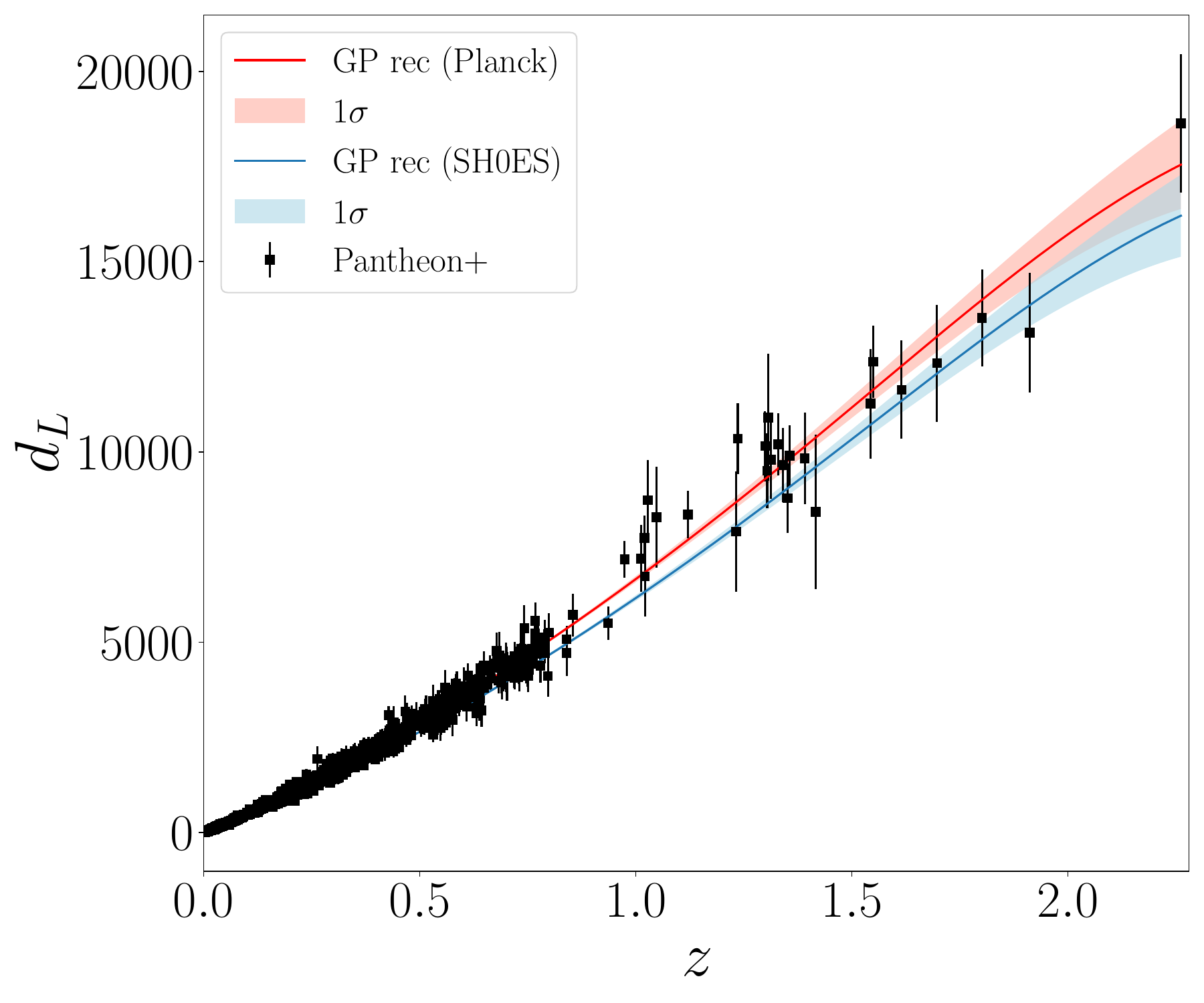}}
{\includegraphics[width=0.32\linewidth]{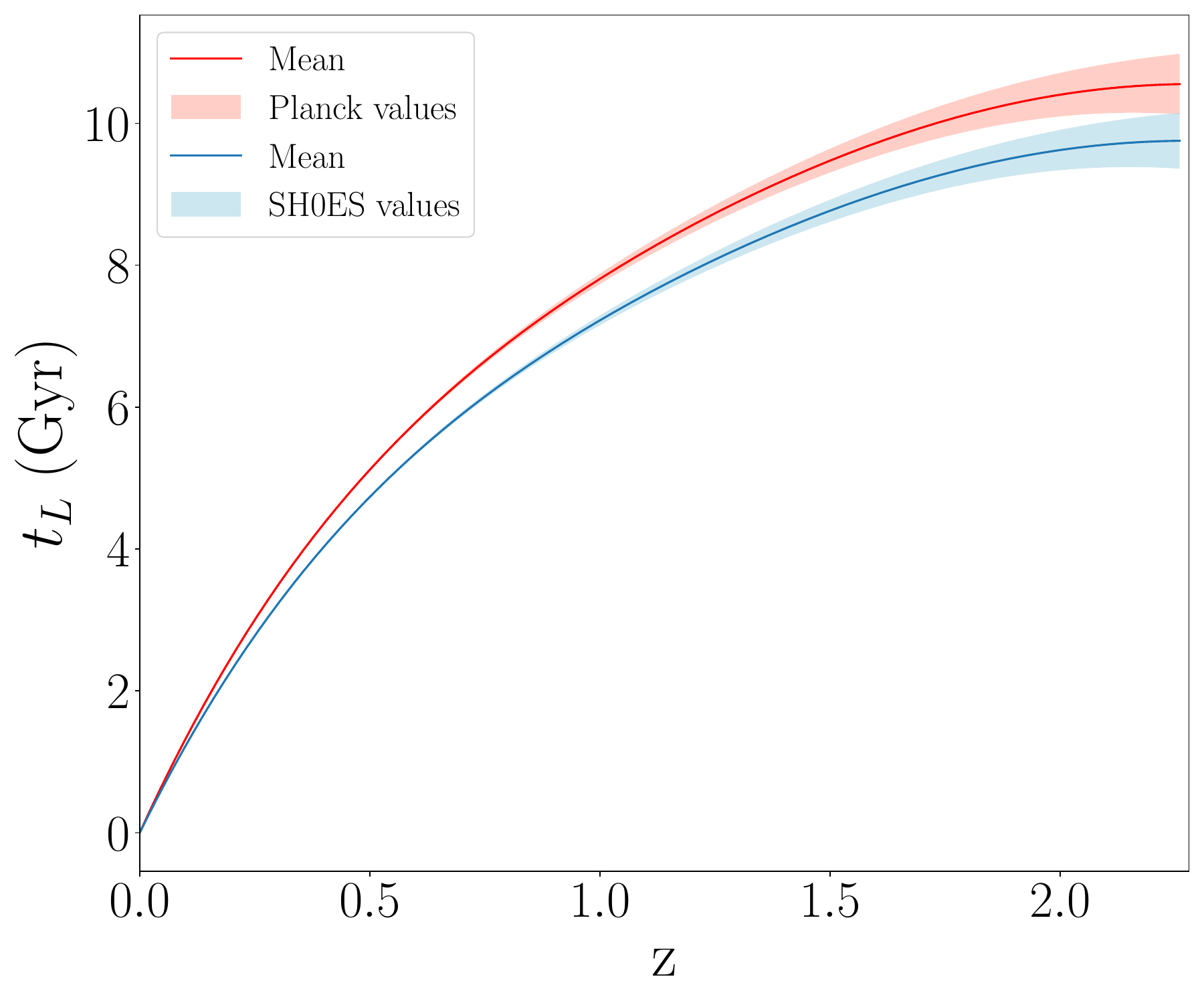}}
{\includegraphics[width=0.32\linewidth]{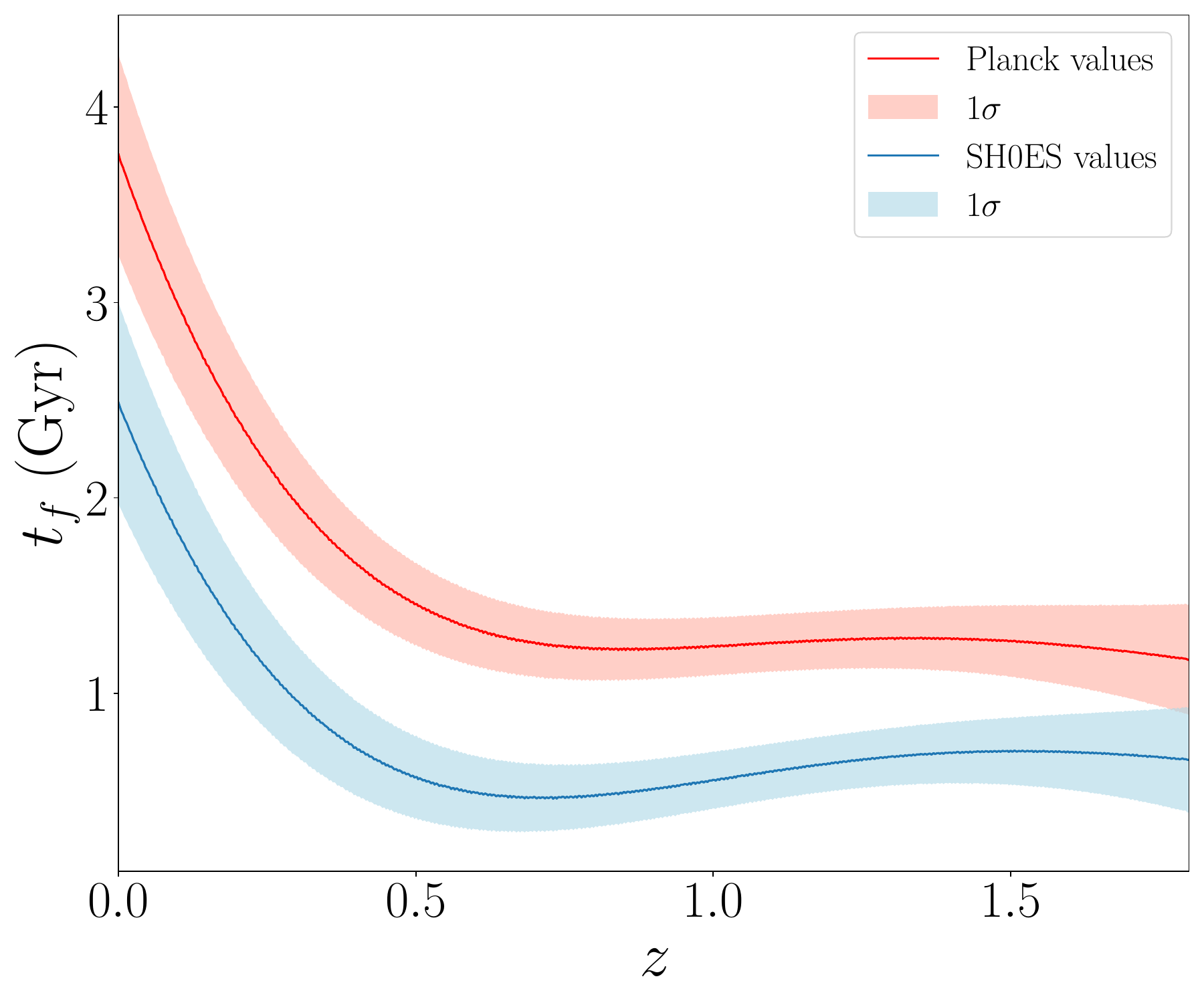}} 
\caption{\textit{Left}: GP reconstruction of the luminosity distance as a function of redshift using the SH0ES and Planck values discussed in the text. \textit{Middle}: The lookback-time redshift relation reconstructed from $d_L(z)$ curves, as expressed by Eq.~\eqref{eq:lookbacktime}. \textit{Right}: Evolution of the galaxy formation time $t_f(z)$ with redshift obtained from Eq.~\eqref{principal}. In all panels, the blue and red regions represent the reconstructed $1\sigma$ intervals assuming SH0ES and Planck parameter values, respectively.}
\label{fig:naoteorico}
\end{figure*}

\subsection{Type Ia Supernova Data}

We use the Pantheon+ catalog \cite{Brout_2022} for the SNe observations, which comprises 1550 SNe within the redshift range $ 0.001 \leq z \leq 2.26$. From the definition of the distance modulus relation,
we obtain the luminosity distance, 
\begin{equation}\label{eq:dl_obs}
d_L^{\text{obs}}(z) = 10^{\frac{m_B(z)- M_B - 25}{5}} \ \text{Mpc}\;,
\end{equation}
where $m_B$ and $M_B$ are the apparent and absolute magnitudes, respectively. We assume two different values for this latter quantity: $-19.42$, corresponding to the $H_0$ value of the Planck Collaboration~\cite{aghanim2020planck} and $-19.253$, corresponding to the measurements of $H_0$ reported by the SH0ES Collaboration~\cite{Riess_2019}. A relevant definition used in the analysis presented in the next section is the theoretical expression for $d_L$, which  in a spatially flat FLRW universe is given by
\begin{equation} \label{dl}
d_L(z) = \frac{c(1+z)}{H_0} \int_{0}^{z} \frac{dz'}{{\mathcal{H}}(z')}\;.
\end{equation}

\section{Results}\label{result}


From Eq.~\eqref{tf}, $t_f$ can be estimated as the difference between the theoretical curve of the age of the Universe and the ages of galaxies. Fig.~\ref{fig:agegapp} shows the GP reconstruction of the ages of 32 galaxies as a function of redshift. The blue contours indicate $1\sigma$ confidence level (C.L.) while the solid black line represents the mean of the reconstruction. The black points correspond to the original data from \cite{simon2005constraints} with their respective uncertainties. The total age of Universe is obtained from Eq.~\eqref{eq:t_0} assuming a flat, $\Lambda$CDM model with the following parameters: (i) $H_0 = 73.04 \pm 1.04$ km s$^{-1}$ Mpc$^{-1}$ and $\Omega_m=0.33$, corresponding to the values from the SH0ES Collaboration -- see the upper curve in left panel of Fig.~\ref{fig:30}; (ii) $H_0=67.4 \pm 0.5$ km s$^{-1}$ Mpc$^{-1}$ and $\Omega_m=0.315$, corresponding to Planck's best-fit scenario -- see the upper curve in the middle panel of Fig.~\ref{fig:30}.  

The right panel of Fig.~\ref{fig:30} shows the evolution of $t_f$ with redshift. In that panel, the upper light red curve corresponds to the Planck's $H_0$ and $\Omega_m$ values, while the lower blue curve corresponds to the SH0ES values for the same parameters. These curves represent $1\sigma$ C.L. intervals for the formation time and show noticeable variations at lower redshifts ($z \lesssim 0.5$). The median value of $t_f(z)$ found for each curve is $\left<t_f\right> = 1.26^{+0.10}_{-0.11}$ Gyr (Planck) and $\left<t_f\right> = 0.72^{+0.14}_{-0.16}$ Gyr (SH0ES).

When we replace the GP reconstruction method with the SR method, we follow the same procedure as before. The results show that although the uncertainty estimates provided by PySR are not as precise as those obtained with GP, the overall behavior remains consistent. The median values of $t_f(z)$  for each curve in the right panel of Fig. 3 are $\langle t_f \rangle = 1.27 \pm {0.25}$  Gyr (Planck) and $\langle t_f \rangle = 0.72 \pm 0.26$ Gyr (SH0ES), indicating that both reconstruction techniques (GP and SR) yield the same behavior for $t_f(z)$.


The second approach uses measurements of the SN luminosity distance to reconstruct the lookback time-redshift relation $t_L(z)$, which is defined as the difference between the total age of the Universe, $t_0$, and its age at a given redshift $z$, $t(z)$. In its integral form, $t_L(z)$ can be expressed as \cite{pires2006lookback}
\begin{equation}\label{lookback}
    t_L(z)=H_0^{-1}\int_{0}^{z}\frac{dz'}{(1+z')\mathcal{H}(z',\textbf{p})}\;,
\end{equation}
which can be integrated by parts, yielding 

\begin{equation}\label{partes}
     t_L(z)=  \frac{1}{1+z}  \left[\int_{0}^{z}\frac{dz'}{H(z')}\right] + 
     \int_{0}^{z}\frac{dz'}{(1+z')^2}\left[\int_{0}^{z'}\frac{dz''}{H(z'')}\right]\;.
\end{equation}
Using the definition of luminosity distance given by Eq.~\eqref{dl}, we obtain
\begin{equation}\label{eq:lookbacktime}
     t_L(z)=\frac{1}{c}\frac{d_L(z)}{(1+z)^2}+ \frac{1}{c}\int_{0}^{z}\frac{d_L(z')dz'}{(1+z')^3}\;.
\end{equation}
From Eq.~\eqref{eq:dl_obs}, we perform a GP reconstruction of $d_L(z)$ assuming $M_B = -19.42$ and $M_B = -19.253$ , which correspond to the values of $H_0$ reported by the Planck and SH0ES collaborations, respectively \cite{vonMarttens:2025dvv} -- see left panel of Fig.~\ref{fig:naoteorico}. Along with the values of $\Omega_m$ presented earlier, we also calculate the total age of the Universe as $t_0 = 13.8$ Gyr (Planck) and $t_0 = 12.43$ Gyr (SH0ES) at 1$\sigma$ level. 

Using the definitions above, the formation time can be expressed in terms of the observed lookback time as \cite{Pires_2006,Santos:2007zza,Dantas:2006dy}
\begin{equation}{\label{principal}}
t_f=t_0-t_g-t_L \;,
\end{equation}
where $t_0$ is the age of the universe at $z = 0$ and the lookback time is calculated at the redshift of a given galaxy $z_g$. Thus, assuming that two cosmic objects at the same redshift possess the same age, we combine the above quantities to calculate $t_f$. The right panel of Fig.~\ref{fig:naoteorico} shows the result of this analysis, where the upper light red contour corresponds to the Planck values ($t_0=13.8$ Gyr) and the lower blue contour corresponds to the SH0ES values ($t_0=12.43$ Gyr). Similarly to the results obtained using the first approach, we also find that $t_f$ varies significantly at lower redshifts. From this analysis, we obtain the following median values for the formation time: $\left<t_f\right> = 1.35_{-0.23}^{+0.21}$ Gyr (Planck) and $\left<t_f\right> = 0.71 \pm {0.19}$ Gyr (SH0ES) at 1$\sigma$ level. For completeness, we also perform the same analysis by replacing the GP reconstruction of $t_L$ with the SR method. Although not shown in Fig.~\ref{fig:naoteorico}, both methods agree with each other and also provide the same  behavior for $t_f$.

Finally, it should be mentioned that our results show significant differences (within $\sim 3\sigma$) in the formation time estimates obtained from SH0ES and Planck values, which is a direct consequence of the well-known tension in the measurements $H_0$ of both collaborations (see e.g. \cite{Freedman:2025nfr} for a recent discussion). In particular, the larger (smaller) the value of the Hubble constant, the smaller (larger) the total age of the universe and the estimated galaxy formation time.

\section{Conclusion}\label{conclu}

Understanding the galaxy formation time ($t_f$) - the cosmic duration needed for a galaxy to assemble the bulk of its stellar mass - is crucial to constraining models of structure formation and the cosmological framework, providing insights into galactic evolution.

In this paper, we employed two approaches to investigate the evolution of $t_f(z)$ from age estimates of 32 high-$z$ galaxies and the Pantheon+ type Ia supernova data. To reconstruct the redshift evolution of galaxy ages and the SN luminosity distance, we used the GP and SR reconstruction techniques. The first approach relies on Eq.~\eqref{eq:t_0} and estimates of the formation time are obtained assuming the $\Lambda$CDM model. Using the GP reconstruction method, we found that the median values of $t_f(z)$ are $\left<t_f(z)\right> = 0.72_{-0.16}^{+0.14}$ Gyr and $\left<t_f(z)\right> = 1.26_{-0.11}^{+0.10}$ Gyr at 1$\sigma$ level for the $H_0$ values provided by the SH0ES and Planck collaborations, respectively. These values are in excellent agreement with those obtained when the GP reconstructions are substituted with the SR ones, with both methods indicating the same behavior regarding the time evolution of $t_f(z)$.

Additionally, without explicitly considering a cosmological model, we also estimated $t_f(z)$ from the reconstruction of the look-back-time redshift relation from SNe data, as described in Eq.~\eqref{principal}. Fixing only the value of $t_0$, we found $\left<t_f\right> = 0.71 \pm 0.19$ Gyr (SH0ES) and $\left<t_f\right> = 1.35_{-0.23}^{+0.21}$ Gyr (Planck) at 1$\sigma$ level. An important aspect worth mentioning is that, regardless of the method used, $t_f$ varies significantly with redshift, a trend that becomes even more pronounced for low-$z$ ($z \lesssim 0.5$). Furthermore, we emphasize that the $t_f(z)$ estimates from both methods presented in this paper are consistent with each other, highlighting the robustness of our analysis.

\begin{acknowledgments}
ASN thanks Coordenação de Aperfeiçoamento de Pessoal de Nível Superior (CAPES) for his grant under which this work was carried out. JCC thanks the financial support from the Brazilian Ministry of Science, Technology, and Innovation through a PCI/ON fellowship. JSA is supported by CNPq grant No. 307683/2022-2 and FAPERJ (Rio de Janeiro Funding Agency) grant No. 259610 (2021). MAD is grateful to the Universidade do Estado do Rio Grande do
Norte (UERN). This work was developed thanks to the High-Performance Computing Center at the National Observatory (CPDON).
\end{acknowledgments}

\appendix
\bibliography{apssamp}
\end{document}